\documentclass[twocolumn]{aastex61}
\pdfoutput=1 
\usepackage{amsmath,amstext}
\usepackage[T1]{fontenc}
\usepackage{apjfonts} 
\usepackage[figure,figure*]{hypcap}

\shorttitle{Dust made simple}
\shortauthors{Gall \& Hjorth}

\newcommand{\msun}{M$_\odot$}
\newcommand{\msunyr}{M$_\odot$ yr$^{-1}$}

\newcommand{\dout}{\bgroup \markoverwith{\rule[0.2ex]{0.1pt}{0.4pt}\rule[0.8ex]{0.1pt}
{0.4pt}}\ULon}

\begin{document}
\title{Maximally dusty star-forming galaxies: Supernova dust production 
and recycling in Local Group and high-redshift galaxies}
\author{C. Gall}
\affiliation{Dark Cosmology Centre, Niels Bohr Institute, 
University of Copenhagen, Juliane Maries Vej 30, DK--2100 Copenhagen, Denmark}
\author{J. Hjorth}
\affiliation{Dark Cosmology Centre, Niels Bohr Institute, 
University of Copenhagen, Juliane Maries Vej 30, DK--2100 Copenhagen, Denmark}
%
\begin{abstract}
Motivated by recent observations suggesting that core-collapse supernovae 
may on average produce $\sim$0.3 \msun\  of dust, we explore a simple dust 
production scenario which applies to star-forming galaxies in the local environment 
(the Magellanic Clouds and possibly the Milky Way) as well as to high redshift (sub-
millimeter, QSO, Lyman break) galaxies. We assume that the net dust destruction (due 
to supernova reverse shock, shocks in the interstellar medium, or astration) is 
negligible on a timescale of 1 Gyr, in which case the dust mass can be estimated as 
0.004 times the star-formation rate (for a Chabrier IMF) multiplied by the duration of 
the star-formation episode. The model can account for observed dust masses over 
four orders of magnitude and across the redshift range 0--8.4, with dust production 
rates spanning five orders of magnitudes. This suggests that star-forming galaxies may 
be seen as maximally dusty, in the sense that a dominant fraction of the dust-forming 
elements forged in a supernova eventually will go into the solid phase. In turn, this 
indicates little destruction of supernova dust or almost complete replenishment, on a 
short time scale, of any dust that is destroyed.
\end{abstract}
\section{Introduction}
\label{s:intro}
A popular picture of cosmic dust production stipulates that asymptotic giant 
branch (AGB) stars are major producers of dust and/or provide the seeds 
for subsequent growth in molecular clouds or the interstellar medium (ISM) 
through accretion to reach the levels of dust observed in most galaxies 
\citep[e.g.,][]{2009ASPC..414..453D,2012MNRAS.422.1263H}. However, recent 
observations challenge this scenario. In particular, the large dust masses 
found in high-redshift QSOs \citep[e.g.,][]
{2008ApJ...687..848W, 2010ApJ...712..942M, 2018Natur.553...51M} cannot be 
accounted for by AGB star dust formation because of the limited time available 
since the Big Bang \citep{2007ApJ...662..927D, 2011A&ARv..19...43G}. In parallel, 
closer to home, \citet{2009MNRAS.396..918M} pointed out a `missing dust-mass 
problem' in the Large Magellanic Cloud (LMC) arising from the inability of existing AGB 
stars to produce the dust mass observed in the ISM. A similar problem exists in the 
Milky Way (MW) \citep{2009ASPC..414..453D} and the Small Magellanic Cloud (SMC) 
\citep{2012ApJ...748...40B,2013MNRAS.429.2527M}. Moreover, the dust emission in 
the LMC and SMC appears to trace the ISM gas rather than the stellar content of the 
galaxies \citep{2013AJ....146...62M}. Thus, in the Local Group, AGB and red 
supergiant stars appear to directly contribute only a maximum of 5--10 \% to the total 
dust budget.

Current literature is surprisingly divided regarding the origin of cosmic 
dust. For example, there are claims that AGB stars 
\citep{2011MNRAS.416.1916V, 2012ApJ...748...40B}, 
supernovae (SNe) \citep{2011A&A...528A..14G, 2011A&ARv..19...43G, 2016MNRAS.463L.112F} or grain growth in molecular clouds 
\citep{2009ASPC..414..453D, 2010A&A...514A..67M,  2010A&A...522A..15M, 2012MNRAS.422.1263H} 
are the prime sources. Fairly advanced chemical evolution models  
\citep[see, e.g.,][]{1998ApJ...501..643D, 2008A&A...479..669C, 2008A&A...479..453Z, 2011A&A...528A..13G, 2011A&A...528A..14G, 2014MNRAS.441.1040R, 2015A&A...577A..80M, 2017MNRAS.471.3152P}
or large scale low resolution cosmological simulations 
\citep{2017MNRAS.468.1505M, 2018MNRAS.tmp.1371A} are often used as tools to 
describe  the complex physics and astrophysics of the life cycle of cosmic dust. The 
differing conclusions reached suggest that the outputs of such models, in part,
reflect the physical prescriptions adopted for the formation, transformation, 
and destruction of dust as well as the input parameters chosen, such as the
assumed net dust productivity of SNe. Furthermore, while the number of measured 
parameters of a galaxy is often limited to a handful (such as the dust mass, the stellar 
mass, the star formation rate, the metallicity, or the dust-to-gas ratio), chemical 
evolution models easily comprise several dozens of unknown or uncertain 
parameters, some of which may potentially strongly impact the final 
conclusions.

In this paper we investigate a simple unifying scenario in which 
core-collapse SNe are hypothesized to be responsible, directly or indirectly, 
for producing the majority of the dust in high-redshift star-forming massive 
galaxies, as well as the SMC, the LMC, and possibly even the MW 
\citep{2016MNRAS.457.3775M}. To set the stage, we discuss the state-of-the-art 
picture of SN dust production and destruction as well as grain growth in the dense ISM 
in Section~\ref{s:review} (this Section can be skipped from a cursory reading). 
We next develop a model involving only few parameters and use recent 
measurements of the amount of dust produced by SNe in Section \ref{s:model}. 
We present existing observational constraints on Local Group galaxies as well as 
high-redshift dusty galaxies in Section \ref{s:data} and discuss the results 
in Section \ref{s:discussion}. 

Our analysis shows that SNe can barely produce all the dust provided that none of the 
newly formed SN dust is destroyed. Alternatively, dust has to reform about as quickly 
as it was formed in the first place and not be prone to subsequent destruction by the 
processes, which destroyed the original SN dust. In this sense, (dusty) galaxies can be 
seen as maximally dusty. This result is illustrated in Figure~\ref{f:md} which relates observed dust masses to those estimated from Equation~\ref{eq:md}
in Section~3.
\section{State of the art: 
SN dust production, destruction and grain growth}
\label{s:review}
Massive stars and core-collapse SNe are known to produce dust 
\citep[e.g.,][see review by \citealt{2011A&ARv..19...43G}]{
1967AnAp...30.1039C,
1970Natur.226...62H,
1979Ap&SS..65..179C,
1989ApJ...344..325K,
1993ApJS...88..477W,
2001MNRAS.325..726T,
2003ApJ...598..785N,
2004MNRAS.352..457P,
2005ApJ...627L.113B,
2008ApJ...680..568S,
2009ApJ...691..650F,
2009ApJ...704..306K,
2010ApJ...713....1C,
2011ApJ...743...73K,
2012ASSL..384..145S,
2013ApJ...776..107S}.
The main issues regarding this scenario are: do core-collapse SNe produce 
sufficient amounts of dust and will what they produce survive the transit 
into the ISM \citep{2011A&ARv..19...43G} and subsequent destruction in the 
ISM \citep{2011A&A...530A..44J, 2015ApJ...803....7S}.
\subsection{SN dust production}
\label{ss:sndust}
\begin{figure}[htp]
\hspace*{-0.5cm}
\includegraphics[width=1.05\hsize]{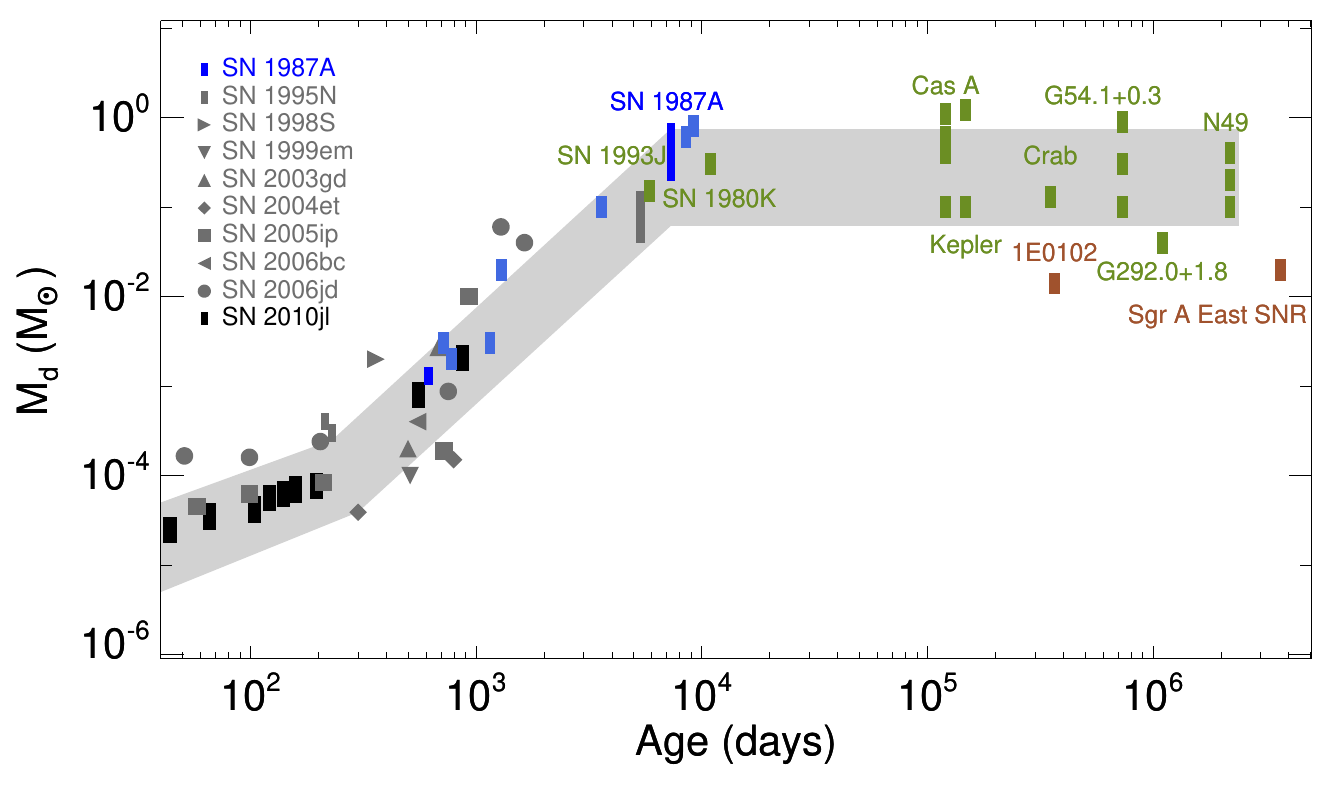}
\caption{Supernova dust mass evolution. SN data indicated as grey and black symbols 
are taken from \citet[][and references therein]{2014Natur.511..326G}. 
Data for SN 1987A \citep{2015ApJ...800...50M, 2012MNRAS.423..451L, 2014ApJ...782L...2I, 2015MNRAS.446.2089W, 2016MNRAS.456.1269B} 
are indicated as blue symbols. For these SNe, the age refers to the time past the 
maximum visual light. SN remnants (SNRs) with cold dust measurements are indicated 
as green bars (age here refers to the time past explosion), namely for SN 1993J and 
SN 1980K \citep{2017MNRAS.465.4044B},
Cas A \citep{2017MNRAS.465.4044B, 2014ApJ...786...55A, 2017MNRAS.465.3309D}, 
Kepler \citep{2009MNRAS.397.1621G},  
Crab \citep{2012ApJ...760...96G, 2015ApJ...801..141O, 2013ApJ...774....8T}, 
G54.1+0.3 \citep{2017ApJ...836..129T, 2017arXiv170708230R}, 
G292.0+1.8 \citep{2016ApJ...831..188G} and
N49 \citep{2010A&A...518L.139O}. 
Red bars mark SNRs with warm dust ($\gtrsim$ 100 K) measurements for 
1E0102 \citep{2009ApJ...700..579R} and Sgr A East SNR \citep{2015Sci...348..413L}.
The grey area between 40 and 7305 days illustrates a possible trend of dust 
formation, while the grey area from 7305 days to 3000 years reflects the
scatter (0.34 \msun\ rms) around the mean value of 0.40 $\pm$ 0.07 \msun\
of the SNRs within this period.  
}
\label{f:devol}
\end{figure}

A discussion of dust masses inferred from SNe is provided by 
\citet{2011A&ARv..19...43G, 2014Natur.511..326G}. While many measurements and 
lower limits are available at early times, few accurate measurements have been made 
at late times, uncontaminated by intervening sources. Figure \ref{f:devol} is a 
compilation of reported dust masses in SNe and SN remnants as a function of time 
since explosion. It is an extended version of Figure 4 in \citet{2014Natur.511..326G}, 
updated with the latest dust mass measurements of SNe and supplemented with 
masses of older SN remnants, including the 10000 yr old SN remnant Sgr A East 
\citep{2015Sci...348..413L}, which has warm ($\gtrsim$ 100 K) dust mass 
measurements. These measurements demonstrate that substantial amounts of dust 
(of order 0.1--1 \msun) can be produced, consistent with most
theoretical estimates \citep{2001MNRAS.325..726T, 2003ApJ...598..785N, 
2010ApJ...713....1C}. Detailed modeling of existing data suggests that dust mass 
either increases over time \citep{2014Natur.511..326G,2015MNRAS.446.2089W}, as
suggested by Figure \ref{f:devol}, or is formed instantly at early times in optically 
thick clumps  \citep{2015ApJ...810...75D}.

Only three SNe, SN~1987A, SN~1054 (Crab), and Cas~A have had their 
dust masses measured accurately. A large dust mass in SN~1987A was first inferred 
by \citet{2011Sci...333.1258M} based on {\it Herschel} data, later confirmed using 
ALMA observations \citep{2014ApJ...782L...2I, 2015ApJ...800...50M}. 
Based on {\it Herschel} and {\it Planck} data, \citet{2012ApJ...760...96G} 
detected significant dust emission from the Crab nebula
\citep[see also][]{2013ApJ...774....8T, 2015ApJ...801..141O, 
2015MNRAS.449.4079M}. Dust has been detected in the Cas A 
\citep{2003Natur.424..285D, 2009MNRAS.394.1307D, 2010A&A...518L.138B} and 
N49 \citep{2010A&A...518L.139O} SN remnants but in these cases it is challenging to 
distinguish dust formed by the SN, interstellar dust, or foreground dust. 
\citet{2017MNRAS.465.3309D} recently managed to overcome this challenge for Cas A.

Core-collapse SNe interacting with their circumstellar material have also been 
shown to be dust producers 
\citep{2008ApJ...680..568S, 2008MNRAS.389..141M, 2009ApJ...704..306K, 
2011ApJ...732..109M, 2014Natur.511..326G}. Finally, massive progenitors of 
SNe, such as luminous blue variable stars, also appear to produce dust with 
a standard dust-to-gas ratio \citep{2012ASSL..384..145S}. For example, 
about 0.4 \msun\ of dust within $2\times 10^{17}$ cm 
has been inferred in the massive erupting star, $\eta$ Car 
\citep{2010MNRAS.401L..48G,2017ApJ...842...79M}.
Such dust is usually assumed to be evaporated by the shock
breakout radiation from the ultimate supernova explosion, but 
dust at sufficiently large radius may possibly survive, depending on the
details of the shock breakout radiation.
%
\subsection{SN reverse shock dust destruction}
\label{ss:snreverse}
The expanding forward shock resulting from the SN explosion sweeps 
up any circumstellar and interstellar material, ultimately leading to 
the creation of a reverse shock \citep[e.g.,][]{1977ARA&A..15..175C}. 
The presence or absence of substantial circumstellar material may influence 
the timescales of when the reverse shock interacts with the SN-formed 
(ejecta) dust. For example, using theoretical models which do not consider 
any circumstellar material, it has been suggested that a reverse shock, 
launched by the swept up interstellar material only, may destroy the majority 
of the dust produced by the SN, on time scales of $\sim$ 10$^{4}$ years 
\citep[for an ambient density $n\sim 0.1$--10 cm$^{-3}$, 
e.g.,][]{2007MNRAS.378..973B}.  
On the other hand, simulations of early reverse shocks in Type IIn SNe 
interacting with a dense circumstellar shell suggest that the reverse shock 
oscillates between layers of swept up material due to radiative cooling
on timescales of tens of days, too short to affect newly formed dust but 
able to explain ripples in early lightcurves \citep{2013ApJ...768..195W}.

In theoretical models \citep{2007MNRAS.378..973B, 2016A&A...590A..65M, 2016A&A...587A.157B, 2016A&A...589A.132B} the properties of the reverse shock 
\citep[e.g.,][]{1977ARA&A..15..175C} are often based on the formalism 
described in \citet[][]{1999ApJS..120..299T} and references therein. 
Rather different conclusions regarding the effect of the reverse shock are reached. For 
example, \citet{2007ApJ...666..955N} find large survivability rates while 
\citet{2007MNRAS.378..973B} find a large destruction rate. These rates 
depend on the assumed destruction mechanisms as well as the morphology of 
the SN remnant and surrounding material. 

Part of the discrepancy can also be traced back to the input model of the dust grain 
size distribution. Thus, models which consider either large grains or dust residing in 
dense ejecta clumps, lead to survival rates in the range 40--98 \% 
\citep{2010ApJ...715.1575S, 2012ApJ...748...12S, 2016A&A...589A.132B, 
2008ApJ...682.1055N}, while, depending on the ambient density,
only $\lesssim$ 2--20 \% survives in models
with very small grains \citep{2007MNRAS.378..973B, 2015MNRAS.451L..70M, 
2016A&A...590A..65M, 2016A&A...587A.157B}. 

Other mechanisms may significantly alter the effects of reverse shock 
interaction with the SN formed dust. ISM magnetic fields may lead to dust 
grains being trapped in the heart of a SN remnant 
\citep{2018arXiv180106859F}. In models taking dust charge and magnetic fields into 
account, dust filaments may be created by non-linear manifestations of highly-
supersonic resonant drag instabilities \citep{2018arXiv180110166H}. Such filaments 
could drift through the reverse shock, thereby possibly avoiding effects of the reverse 
shock entirely. From an observational perspective, the detection of about 0.02 \msun\ 
of warm dust ($\sim$ 100 K) in the $\sim$ 10000 year old SN remnant Sagittarius A 
East in the Galactic center suggests that there is limited dust destruction due to the
reverse shock \citep{2015Sci...348..413L}.

Reverse shocks in SNe and SNRs have also been detected on much shorter 
timescales, i.e., of order tens or hundreds of years. SN 1998S exhibited reverse shock
emission after 14 years, at which point significant amounts of dust
was reported in the unshocked ejecta inside the reverse shock as well
as in the CSM, outside the reverse shock \citep{2012MNRAS.424.2659M}.
Ongoing reverse shock activity has also been detected in the outer shell of 
SN 1987A \citep{1998ApJ...492L.139S, 2011ApJ...743..186F, 2013ApJ...768...88F}, 
in Cas A \citep{2004ApJ...614..727M, 2013A&A...558L...2W} and in SN 1006 
\citep{2011ApJ...742...80W}. There are also indications of ejecta dust in other 
decades-old SNe \citep{2012ApJ...751...25M}, although it is unclear to what extent 
the remnants have been affected by a reverse shock. On the other hand, so far no 
such activity has been seen in the $\sim$ 1000 year old SN 1054 (the Crab nebula), 
where the dust seems to be shielded in dense filaments and well set to begin 
its transit into the ISM \citep{2012ApJ...760...96G}. 

In summary, it is at present unclear to what extent the SN produced dust
will survive the destruction by the various short and long time-scale 
reverse shocks. 
\subsection{SN forward shock ISM dust destruction}
\label{ss:snforward}
A SN forward shock can strongly affect the existing dust in the CSM or ISM. 
A variety of physical processes are at work, ranging from shattering, which 
affects the grain size distribution, to destructive processes, such as sputtering 
or vaporization, which decrease the total dust mass 
\citep[e.g.,][]{1978MNRAS.183..367B, 1996ApJ...469..740J, 2004ApJ...614..796S}. 
These destruction processes apply to any kind of dust, not just dust produced by SNe.

Ongoing dust destruction of swept up ISM dust grains has been seen in 
the SNRs N49 \citep{2016ApJ...826..150D} and Puppis A 
\citep{2010ApJ...725..585A}. The observations suggest that about 30--90 \% of the
Fe-bearing grains around N49 and about 25\% of the graphite and silicate dust 
around Puppis A is destroyed. About 35\% of the mass of grains are destroyed 
inside a 0.14 pc region behind the forward shock of the Cygnus Loop SNR \citep{2010ApJ...712.1092S}.

\citet{2015ApJ...799...50L} and \citet{2015ApJ...799..158T} found very large 
destruction rates of SNe from studies of SN remnants in the Magellanic 
Clouds. They concluded that there is less dust inside the SN remnant, 
implying a very large destruction rate and a very short dust survival time.\footnote{We note that the formal significance of the 
\citet{2015ApJ...799...50L} results is marginal and may be affected 
by warm dust preventing any cold dust present to be detected 
\citep{2018SSRv..214...53M}. The short dust lifetimes obtained by 
\citet{2015ApJ...799..158T} are based on the formalism of 
\citet{2007ApJ...662..927D} to calculate SN dust destruction efficiencies 
as a function of SN shock velocity. We note that the analysis assumes
small grains and 100 \% destruction at shock velocities of 500 km/s, at variance 
with the observations of Puppis A suggesting destruction of up to 
25 \% of the grains at such velocities \citep{2010ApJ...725..585A}.} 
These works suggest that SNe destroy more dust than they form, 
and hence would imply that SNe should in principle destroy all dust on very 
short time scales, 20--50 Myr. Destruction time scales shorter than 
injection time scales (60 Myr) for Si-bearing grains in the ISM were also 
found by \citet{1998ApJ...499..267T}. Other estimated time scales are longer, of order 
100 Myr \citep{2011A&A...530A..44J} or 1000 Myr for the SMC \citep{2013MNRAS.429.2527M}. 

Very short destruction time scales could imply that dust would not be present 
in most galaxies. It has therefore been suggested that SNe cannot be allowed to 
be efficient dust destroyers if we are to observe the dust we see in galaxies 
\citep{2011A&A...528A..13G,2011A&A...528A..14G,2014ApJ...782L..23H}. As in the 
case of the reverse shock models, the dust composition and grain size 
distribution as well as the morphology of the ISM are important. Recent estimates 
of ISM dust destruction time scales have gone  up significantly, to 
$\sim$ 2--3 Gyr for (small grain) silicates \citep{2015ApJ...803....7S}. 
These models consider that SN dust destruction is efficient only in the warm 
phase of the ISM, which in the work by \citet{2015ApJ...803....7S} is assumed 
to make up 80\% of the ISM. The time scales increase because dust 
destruction in the cold and hot phases of the ISM may be inefficient 
\citep{2015ApJ...803....7S} and are also enhanced for larger grains or 
a lower filling factor of the warm ISM as larger grains are more robust 
\citep[e.g.,][]{2010ApJ...715.1575S}. 

Clustered SN explosions, as opposed to isolated SN explosions, can cause a net 
increase in the amount of dust in the surroundings of young massive stellar clusters 
\citep{2018arXiv180806614M}. At small SN shock destruction rates (i.e., for masses 
of ISM gas cleared of dust by a single supernova remnant of less than about 
100 \msun), astration will dominate the dust destruction in galaxies \citep{2014ApJ...782L..23H}.
%
\subsection{Grain growth in the dense ISM}
\label{ss:thesavior}
%
Dust grains can grow through accretion of ions in the ISM such as in dense molecular 
clouds or in the cold neutral medium \citep[e.g.,][]{2009ASPC..414..453D}.
This process is often invoked in the Milky Way 
\citep{2009ASPC..414..453D, 2012MNRAS.422.1263H, 2016ApJ...831..147Z}
and more generally at high redshift \citep[e.g.,][]{2010ApJ...712..942M}, 
especially in scenarios in which SNe are assumed to be net destroyers of
dust because of catastrophic reverse shock destruction and forward shock destruction 
of pre-existing ISM dust. Simple grain growth models often consider spherical grains 
and assume that all accreted atoms stick onto the grains, thereby increasing the mass 
of grains, possibly yielding rapid accretion time scales of $\lesssim$ 1 Myr \citep[e.g.,][]
{2012MNRAS.422.1263H}, depending sensitively on the density of the 
medium, the type of ion, and the sizes of the seed grains.

Chemical evolution models typically treat grain growth in a parametrized way primarily 
determined by the amount of gas phase metals instantly mixed into an 
estimated number of clouds with assumed physical conditions (lifetimes, 
temperatures, densities) similar to those in the Milky Way \citep[e.g.,][]
{2009ASPC..414..453D, 2011MNRAS.416.1916V, 2011A&A...525A..61P, 
2013EP&S...65..213A, 2014MNRAS.438.2765C, 2015MNRAS.451L..70M, 
2017MNRAS.471.3152P, 2018MNRAS.473.4538G}.    

While grain growth in cold (a few tens of K) dense molecular clouds in the 
Milky Way is an important (but slow) process, the physical conditions in 
massive starburst galaxies is likely far from similar to those in the Milky 
Way \citep{2018MNRAS.476.1371C}. Indeed, \citet{2016MNRAS.463L.112F} 
point out that at high redshift the higher CMB temperature 
(T$_{CMB}$=2.725(1+$z$)) prevents dust to cool to temperatures 
($\lesssim$ 20--30 K) required for particles to stick to the surfaces of seed 
grains, hampering efficient grain growth. Additionally, dust growth in cold molecular 
clouds largely proceeds through the coating of refractory seed grains with icy mantles 
composed of primarily water ice and some molecules such as, 
e.g., NH$_{3}$, CO, CO$_{2}$ and more complex ones 
\citep{2012A&ARv..20...56C,2015ARA&A..53..541B} 
through hydrogenation and oxidation of atoms and simple molecules on the 
surfaces of grains. Dust forming elements such as, e.g., Si, or C are at best 
embedded as impurities. While ice mantles possibly foster coagulation 
processes in cold molecular clouds \citep{2013ApJ...777...73B} on times 
scales of 10$^7$ yr \citep{2009A&A...502..845O}, rapid and efficient grain 
growth through accretion processes, however, is uncertain. For example, 
\citet{2018MNRAS.476.1371C} found that grain growth of Si-bearing grains 
through accretion of elemental Si remains impossible at any redshift. 

Additionally, the lifetimes of giant molecular clouds is short, of order 
4--25 Myr \citep{2013MNRAS.432..653D, 2017MNRAS.464..246B} and smaller dense 
molecular clumps, such as the birth places of massive stars in the LMC, may 
only live for less then 1 Myr \citep{2012ApJ...751...42S}. 
The mixing time scales may also be longer than `instant', up to 125--350 Myr for 
massive galaxies \citep{2002ApJ...581.1047D, 2018MNRAS.475.2236K}. Longer 
mixing timescales may cause an imbalance between metal production and diffusion, 
which possibly explains the large diversity of metallicity distributions in 
high redshift galaxies \citep{2015MNRAS.449.2588P}. 

On the other hand, metal depletion studies show that up to 99\% of all the iron 
ejected by core collapse SNe or SNe Ia is depleted in the various temperature phases 
of the ISM of local as well as higher redshift galaxies 
\citep{2009ApJ...700.1299J, 2016A&A...596A..97D}. 
Elemental depletion patterns have been suggested to be a signature of grain growth in 
the ISM through accretion of ions onto pre-existing grains \citep{2016ApJ...825..136D, 
2016A&A...596A..97D,2018ApJ...857...94Z}. 
The large depletion rates of iron combined with the lack of evidence
for dust formation by SNe Ia
\citep[e.g.,][and references therein]{2011A&ARv..19...43G,2018ApJ...856L..24N},     
which produce a dominant fraction of the iron in the universe, suggests that a
substantial fraction of the iron must be incorporated in dust grains
grown in the ISM. Conversely, the detection of crystalline silicates are
hard to reconcile with dust grown in the ISM \citep{2016ApJ...825..136D}.
%
\section{Estimating the dust mass due to supernovae in a star-forming galaxy}
\label{s:model}
%
In this section we develop a simple model for the dust mass in a galaxy
with a given (simple) star-formation history.
For a star-forming galaxy, we estimate the total dust mass produced by SNe by making the following assumptions:
\begin{itemize}
\item The dust is produced during a star-formation episode of duration 
$\Delta t$ with a mean star-formation rate $\psi$.
\item The life times of dust-producing massive stars are negligible compared to $\Delta t$.
\item The rate of SNe (i.e., of massive stars) is proportional to the
star-formation rate, i.e., $\gamma\psi$. 
\item Massive stars (SNe and their progenitors) on average produce $\eta$ 
solar masses of dust. 
\item On time scales shorter than $\Delta t$, the net growth or destruction 
of dust in the ISM is negligible.
\end{itemize}

Under these assumptions the resulting dust mass is proportional to the average star-formation rate during the star-formation episode,
\begin{equation}
M_d=\mu_D \psi \Delta t ,
\label{eq:md}
\end{equation}
with 
\begin{equation}
\mu_D\equiv \gamma \eta,
\end{equation}
\citep[see also][whose notation we adopt in this paper]{2011A&ARv..19...43G,2014ApJ...782L..23H}.  
With this definition, the dust production rate is simply 
\begin{equation}
DPR\equiv\frac{M_d}{\Delta t} =\mu_D \psi.
\end{equation}
In the remainder of this Section we discuss the values and uncertainties of 
the parameters entering these expressions.
\subsection{The proportionality factor between star-formation rate and SN rate, $\gamma$}
\label{ss:gamma}
The number of core-collapse SNe produced per gas mass available for
star formation is
\begin{equation}
\gamma \equiv
\frac{\int_{8~\mathrm{M}_\odot}^{40~\mathrm{M}_\odot} \phi (m) dm} 
{\int_{m_{\rm min}}^{100~\mathrm{M}_\odot} \phi(m) m dm}.
\label{e:gamma}
\end{equation}
Here, the stellar initial mass function (IMF), $\phi(m)$, is defined as the 
number of stars, $\phi(m) dm$, in the mass interval $m$ to $m + dm$, 
where $m$ is the zero age main sequence mass. 
The IMF directly determines the number of massive stars exploding as SNe 
relative to the total mass of stars. It is defined over the range 
$m_{\rm min} <m<$ 100 \msun, with $0.1~\mathrm{M}_\odot \le m_{\rm min} 
< 8~\mathrm{M}_\odot$. The mass range for core-collapse SN formation is 
traditionally taken to be 8 \msun\ $<m<$ 40 \msun\ 
\citep{2003ApJ...591..288H,2013ApJ...765L..43I}. 
The Salpeter IMF, which takes a power-law form,
\begin{equation}
\phi(m) \propto m^{-\alpha},
\label{e:imf}
\end{equation}
with $\alpha = 2.35$, results in $\gamma =  0.007$ \msun$^{-1}$.

It is known that the total SN dust production depends on the IMF 
\citep{2007ApJ...662..927D, 2011A&ARv..19...43G, 2014ApJ...782L..23H},
in particular through $\gamma$. However, so does the total inferred 
star-formation rate. As long as the star formation rate is inferred 
consistently with the same IMF as that used to infer $\gamma$ then our 
results will be largely independent of the adopted IMF because we will 
correctly infer the number of massive stars (and hence dust producing SNe;
note however that a starburst can be sustained for longer for an IMF 
disfavouring low-mass stars). Therefore, in this paper we assume a Salpeter 
IMF and convert reported Chabrier IMF based values to Salpeter IMF values using a 
conversion factor of 1.8. IMF-independent massive star-formation rates 
can be obtained by dividing the Salpeter IMF SFR by 10.

\citet{2015ApJ...813...93S} compared measured volumetric core-collapse SN rates 
to the observed cosmic SFR density 
\citep[for a Salpeter IMF][]{2014ARA&A..52..415M}. The inferred value of $\gamma =  0.0091\pm0.0017$ 
\msun$^{-1}$ suggests that we adopt a 0.15 dex uncertainty in $\gamma$ for a given
IMF.
%
\subsection{The average dust mass per core-collapse SN, $\eta$}
\label{ss:eta}
In Figure~\ref{f:devol} we have updated the evolutionary trend suggested 
by \citet{2014Natur.511..326G} with recent SN dust mass measurements 
including SN remnants. It appears that the maximum amount of dust formed 
in a SN is reached within about 30 years and is not significantly altered 
for another 1000 years. Formally, the mean value reached at late times
is 0.40 $\pm$ 0.07 \msun.

However, this value is based on a very heterogeneous set of reported
measurements and individual systematic uncertainties are hard to quantify. To
obtain a more reliable estimate, we here focus on the three systems alluded to
in Section~2.1, which have exquisite measurements, detailed modeling, and
systematic errors under control. 
We adopt a dust mass of $0.6 \pm 0.3$ \msun\ for SNe 1987A. In the case of 
SN 1054 a dust mass of 0.15 $\pm$ 0.12 \msun\ is consistent 
with all reported values (see Section \ref{ss:sndust} for references). 
Finally, \citet{2017MNRAS.465.3309D} recently obtained a dust mass of
0.2--0.6 \msun\ in Cas A. The weighted mean of these three accurate observed 
dust mass yields is $0.26 \pm 0.10$ \msun\ while the straight mean is 
0.38 \msun.

Such direct averages are likely to be biased since the three SNe may 
not be representative of the population of dust-producing core-collapse SNe. 
To compute an IMF-averaged mean dust mass we can compare to theoretical 
expectations and thereby take into account the distribution of progenitor 
masses.

\begin{figure}[htp]
\hspace*{-1cm}
\includegraphics[width=1.15\hsize]{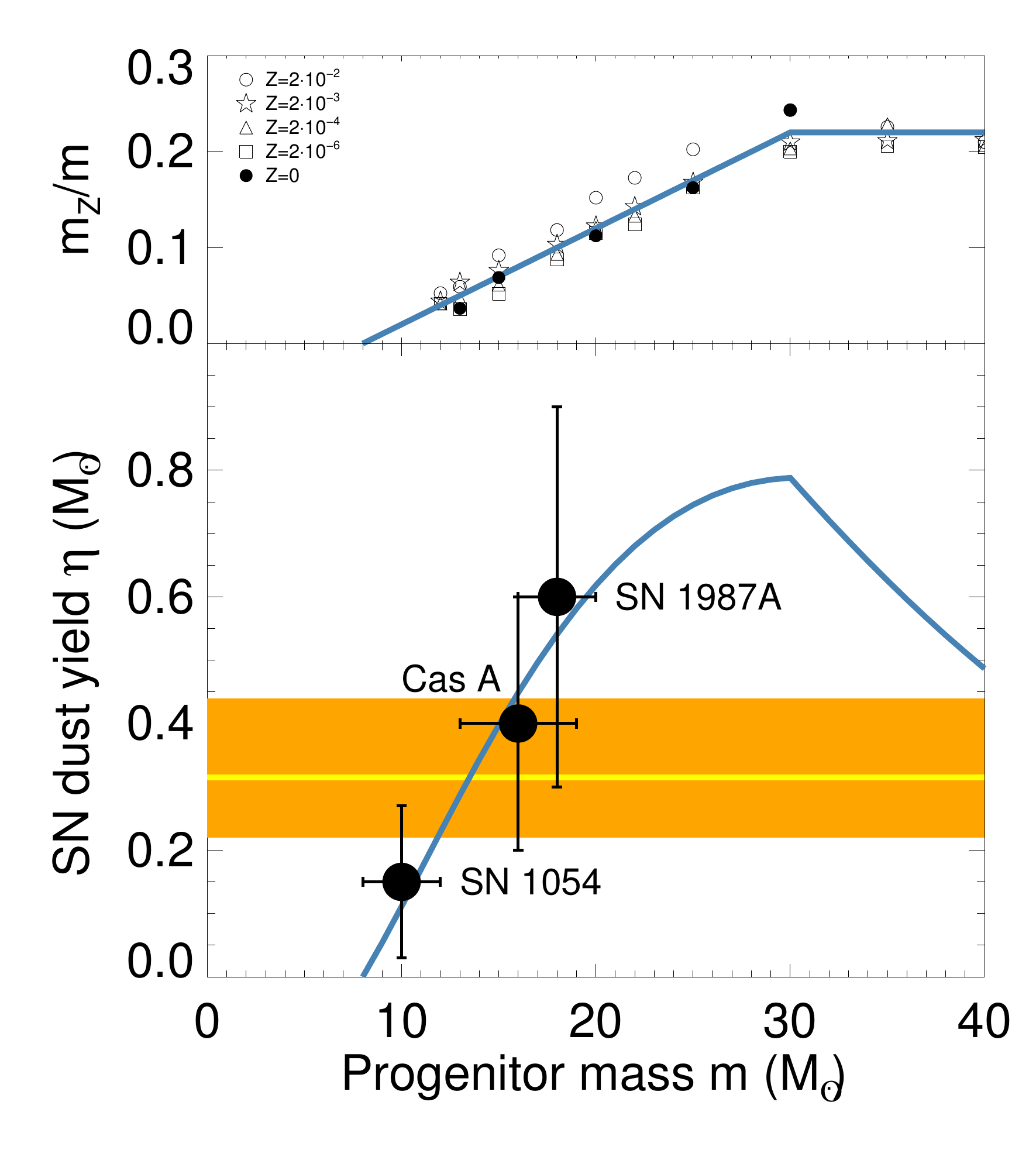}
\caption{
Lower panel: SN dust yield as a function of SN progenitor mass. The observed cold dust masses for the Crab \citep{2012ApJ...760...96G,2013ApJ...774....8T},
SN 1987A \citep{2011Sci...333.1258M,2014ApJ...782L...2I}, and 
Cas A \citep{2017MNRAS.465.3309D} are plotted as filled black circles. The blue curve is a theoretical estimate of the SN dust yield (assuming no destruction) (adapted from Fig.~6 of \citealt{2011A&ARv..19...43G} using Equation 6). 
The IMF averaged dust yield is indicated as the yellow line for a Salpeter IMF. 
The orange band indicates the associated 0.15 dex error adopted. The average value is not strongly sensitive to the uncertain dust yield above 30 \msun\ due to the dominance of stars with lower masses. 
Upper panel:
Relative metal production as a function of progenitor mass for different 
metallicities from \citet{1995ApJS..101..181W} (open symbols) 
and \citet{2002ApJ...565..385U} (filled circle). The blue line
is a representation of these points, as expressed in
Equation (8).
}
\label{f:eta}
\end{figure}

We define the dust yield per SN as
\begin{equation}
\eta\equiv
\frac{\int_{8~\mathrm{M}_\odot}^{40~\mathrm{M}_\odot} \epsilon (m) m_Z(m) \phi(m) dm} 
{\int_{8~\mathrm{M}_\odot}^{40~\mathrm{M}_\odot} \phi(m) dm}. 
\end{equation}
We use the analytical `maximum efficiency' of
\citet{2011A&ARv..19...43G}, 
\begin{equation}
\epsilon (m) \equiv \frac{m_{\mathrm{d}}}{m_Z} = 1.2\exp \left ( - \frac{m}{13 \rm M_\odot} \right ),
\label{e:epsilon}
\end{equation}
which is an empirical expression derived from
theoretical predictions of the dust mass per supernova
\citep{2001MNRAS.325..726T,2003ApJ...598..785N} which does
not account for the effects of reverse shock  dust 
destruction \citep[e.g.,][]{2007MNRAS.378..973B}.
Guided by the nucleosynthetic yields found by \citet{1995ApJS..101..181W} and \citet{2002ApJ...565..385U}
we assume the following expression for the metal production as a function
of progenitor mass (see Figure~\ref{f:eta}),
\begin{equation}
m_Z (m) =
\begin{cases}
0.01 \left ( \frac{m}{{\rm M}_\odot}-8\right ) m ; & 8 <\frac{m}{{\rm M}_\odot}< 30\\
0.22 m ; &  30 \le \frac{m}{{\rm M}_\odot}< 40 .
\end{cases}
\label{e:ww95}
\end{equation}

With these expressions, we obtain an approximately linear
relation between $\eta$ and $\alpha$,
\begin{equation}
\eta/{\rm M_\odot}\approx0.314-0.106~(\alpha-2.35),
\label{e:eta}
\end{equation}
independent of $m_{\rm min}$ (see Figure~\ref{f:eta}).
%
We note the excellent agreement between the observed and model dust yields
depicted in Figure 2. We can therefore infer the expected IMF-averaged value 
of the dust yield per supernova to be 0.31 $M_\odot$ to which we associate an 
uncertainty of 0.15 dex, see Figure 2. Combining the adopted values of $\gamma$ and 
$\eta$ we infer a dust productivity of $\mu_D=0.0022$ \citep{2011A&ARv..19...43G, 
2014ApJ...782L..23H} for a Salpeter IMF or $\mu_D=0.0040$ for a Chabrier IMF, with 
a $\sim 0.2$ dex uncertainty. 
%
\subsection{The star-formation rate $\psi$}
\label{ss:psi}
Observational SFR indicators are sensitive to possible dust extinction if 
inferred from the ultraviolet flux. Moreover, they strongly depend on the 
adopted IMF since the far-infrared or ultraviolet luminosity measures the 
massive SFR, while the extrapolation to total SFR depends on the relative 
number of lower-mass stars. 

We adopt a 0.15 dex uncertainty in $\psi$ for a given IMF, i.e., 
in the massive 
star-formation rate. We quote SFRs converted to Salpeter IMF and use a
corresponding $\gamma$ in computing SN rates. This way the results are
independent of $m_{\rm min}$.
\subsection{The duration of the star-formation episode $\Delta t$}
\label{ss:deltat}
Local Group galaxies have an extended star-formation history during which
there will be time for AGB stars to contribute significantly to the dust 
mass budget \citep{2014MNRAS.442.1440S}. There will also be growth of dust 
grains through accretion in the ISM. Conversely, 
destruction, astration and expulsion will counter these effects.
Thus, $\Delta t$ should be taken to be smaller than the time scales 
relevant to the latter processes. For the SMC, LMC and MW, 
\citet{2013MNRAS.429.2527M} estimated dust destruction time scales of 
1.4 Gyr, 400 Myr and 500 Myr, respectively. These however were based on 
rather high estimates of the SN rates. \citet{2011A&A...530A..44J} estimate 
dust lifetimes of 0.1--1 Gyr in the ISM of the MW. We adopt a fiducial value 
at the upper end of this range, $\Delta t=$ 1 Gyr, with a 0.3 dex
uncertainty, whilst noting that dust destruction time scales are subject 
to current debate, see Section \ref{s:discussion}.

For massive starbursts at high redshift (e.g., sub-mm galaxies) it is likely 
that a large fraction of the dust observed is formed in the current burst. 
For starburst galaxies, the duration can be assessed from the observational
data at hand and typically turns out to be of order 100 Myr. This is 
consistent with recent simulations which suggest that typical starbursts in 
galaxies at $z\sim 6$ have a duration of about 100 Myr 
\citep{2018arXiv180110382C}, with more massive galaxies having slightly 
larger burst durations. In this paper we adopt $\Delta t = 100$ Myr for 
$z>6$ galaxies while we adopt the slightly longer depletion time scale for 
sub-millimeter galaxies (SMGs) at $z\sim 3$.

In total, the combined uncertainty in our theoretical SN dust mass estimates is
$\sim 0.4$ dex.
\subsection{Observational dust mass estimates}
\label{ss:md}
There are significant uncertainties related to determining dust masses 
\citep{2011A&ARv..19...43G}. Far-infrared emission is usually interpreted as 
thermal dust emission. The temperature inferred from modified black body fits 
to the data is degenerate with the dust mass and significant amounts of cold 
dust could in principle be hidden and still be consistent with observational 
constraints. Other issues relate to the assumed type of dust, and hence the 
normalisation of the dust absorption coefficient $\kappa$ and the emissivity 
index $\beta$ \citep[see, e.g.,][for a discussion of observational 
uncertainties related to determining dust masses in SMGs and QSOs]
{2010A&A...514A..67M}.
We adopt an uncertainty of 0.4 dex in observed dust masses which 
matches the uncertainty in the dust mass inferred from the simple model.
%
\section{Data}
\label{s:data}
\begin{deluxetable}{lllll}
\tablecolumns{13}
\tablewidth{0pc}
\tablecaption{Star Formation and Dust Properties}
\tablehead{
\colhead{Galaxy}          &
\colhead{Redshift}        &
\colhead{$\psi$}          &
\colhead{$M_d$}           &  
\colhead{$\Delta t$}      \\
\colhead{}                &
\colhead{}                &
\colhead{(\msunyr)}       &
\colhead{($10^6$ \msun)}  &
\colhead{(Myr)}           \\
}
\startdata
SMC           & 0        &    0.07 &    0.11 & 1000  \\
LMC           & 0        &    0.2  &    1.1  & 1000  \\
Milky Way     & 0        &    3.4  &   25    & 1000  \\
\hline
SMGs          & $\sim 2$--3 &  700    &  400    &  142  \\
\hline
G09 83808     & 6.03     &  640    &  190    &  100  \\
HFLS3         & 6.34     & 1190    &  300    &  100  \\
QSO J1148+525 & 6.42     & 3000    &  590    &  100  \\
SPT0311 E     & 6.90     &  972    &  400    &  100  \\
SPT0311 W     & 6.90     & 5220    & 2500    &  100  \\
\hline
A1689-zD1     & 7.5      &   49    &   50    &  100  \\
MACS0416\_Y1  & 8.31     &   25    &    4    &  180  \\
A2744$\_$YD4  & 8.38     &   37    &    5.5  &  100  \\
\enddata
\tablecomments{The star-formation rates are inferred assuming a 
Salpeter IMF with $m_{\rm min} = 0.1~\mathrm{M}_\odot$. 
The horizontal lines delineate different
object classes, namely Local Group galaxies, intermediate
redshift SMGs, high-redshift SMGs, and high-redshift Lyman-break
galaxies.
These are plotted with different colors in Figure 3.
For notes on individual objects, see Section~\ref{s:data}.
}
\label{tab:data}
\end{deluxetable}
For the purposes of illustrating the range of validity of our simple model,
we here discuss the properties of the most prominent Local Group galaxies as 
well as a selection of representative high-redshift dusty galaxies. 
Table~\ref{tab:data} summarizes the observational estimates of the 
star-formation rate and dust masses, as well as our adopted durations 
of the star-formation episode. The individual objects/object classes are discussed below.

The estimated current star-formation rate of the SMC is 
$3.7\times 10^{-2}$ \msunyr\ \citep{2011ApJ...741...12B}, consistent with 
0.03 \msunyr\ found by \citet{2012ApJ...761...42S}. The current value, 
however, is not directly relevant for our purposes because dust will have 
been built up over a longer period. \citet{2004AJ....127.1531H} inferred a 
burst of star formation lasting 200 Myr, about 400 Myr ago, at 0.25 \msunyr. 
In a more recent study,\footnote{Adopting a Salpeter IMF (D. Zaritsky, 
private communication, 2014).}

\citet{2009AJ....138.1243H} find an average 0.18 \msunyr\ over 100 Myr 
(ranging from 0.1--0.3 \msunyr), while the average over 10 Gyr is around 
0.12 \msunyr. Their high inferred current SFR of order $\sim 0.1$ \msunyr\ suggests 
that we should use the lower boundary of the \citet{2009AJ....138.1243H} study. 
The average lower boundary over 200 Myr (left panel of their Fig.~19) is 
about 0.05 \msunyr\ while it is about 0.07 \msunyr\ over the past 1 Gyr, 
which we adopt. We note that this could be a lower limit.
\citet{2010A&A...523A..20B} estimated a dust mass of 0.29--$1.1 
\times 10^{6}$ \msun\ from the measured sub-mm to cm excess. This could even 
be a lower limit if there is very cold dust. However, 
\citet{2012ApJ...761...42S} (and references therein) argue that such a large 
dust mass would be inconsistent with the amount of metals available and 
instead quote a dust mass of $1.1 \times 10^{5}$ M$_\odot$, which we adopt 
(\citealt{2007ApJ...658.1027L} find a dust mass of $3 \times 10^{5}$ 
M$_\odot$).

For the LMC, \citet{2009AJ....138.1243H} find that star formation has proceeded at an 
average rate of roughly 0.2 \msunyr\ for the past 5 Gyr. \citet{2012ApJ...761...42S} find 
a dust mass of $1.1\times 10^{6}$ \msun, which appears consistent with 
\citet{2009MNRAS.396..918M} who quote a dust mass of $1.6\times 10^{6}$ 
\msun, based on the gas mass and an assumed dust-to-gas mass ratio.

For the MW, \citet{2011AJ....142..197C} find $\psi =1.9\pm 0.4$ \msunyr, 
calibrated to the Kroupa IMF. Correcting to Salpeter IMF (same correction factor as for 
the Chabrier IMF) gives $3.4 \pm 0.7$ \msunyr. \citet{2009ASPC..414..453D} 
estimates the total amount of dust in the ISM as 0.005 times the mass in the ISM of 
$5 \times 10^{9}$ \msun, i.e., $M_d = 2.5 \times 10^{7}$ \msun.

SMGs and high-redshift QSOs typically have star-formation rates of order 
1000 \msunyr\ and dust masses in the range $10^{8-9}$ \msun\ 
\citep{2010A&A...514A..67M,2010A&A...522A..15M,2011A&A...528A..14G,
2018Natur.553...51M}. For a Salpeter IMF we adopt the average values of 700 
\msunyr\ and $4\times 10^{8}$ \msun\ for SMGs up to redshift 4 reported by 
\citet{2010A&A...514A..67M}. The typical duration of such starbursts can be 
estimated from the available gas mass available which is of order $10^{11}$ 
\msun\ \citep{2012A&A...541A..85M}, i.e., 142 Myr. 

A compilation of very high-redshift galaxies was recently provided by
\citet{2018Natur.553...51M}. Below we summarize the observations of galaxies
for which sufficient information is available for our purposes.

For the $z=6.42$ QSO SDSS J1148+525, which has been used as a benchmark in 
many studies of dust formation at high redshift \citep{2007ApJ...662..927D, 
2011A&A...528A..14G, 2011MNRAS.416.1916V}, the estimated SFR is 3000 
\msunyr\ (for a Salpeter IMF) and the dust mass is $5.9\pm 0.7 \times 10^8$ 
\msun\ \citep{2010A&A...522A..15M, 2011A&A...528A..14G}. 
\citet{2011A&A...528A..14G} quote a starburst age of 30 Myr.

The lensed high-redshift ($z=6.34$) SMG HFLS3 \citep{2014ApJ...790...40C}
has an average SFR over 100 Myr of 660 \msunyr\ for a Chabrier IMF, 
corresponding to 1190 \msunyr\ for a Salpeter IMF. The measured dust mass is 
$3 \times 10^8$ \msun.

\citet{2018Natur.553...51M} observed the lensed system SPT0311 at
$z=6.90$, which consists of two separate components, E and W.
The respective star-formation rates (Chabrier) are $540\pm175$ \msunyr\ and 
$2900\pm1800$ \msunyr, while the inferred dust masses are
$0.4\pm0.2 \times 10^9$ \msun and $2.5\pm1.6 \times 10^9$ \msun. The gas 
depletion rates are 74 Myr and 93 Myr.

\citet{2018NatAs...2...56Z} observed the $z=6.03$ lensed dusty galaxy
G09 83808. The Salpeter star-formation rate is $640\pm90$ \msunyr\ and
the inferred dust mass, based on Herschel observations is $1.9\pm0.4
\times 10^8$ \msun. The gas depletion rate is 40 Myr for a Chabrier IMF. 

The three highest redshift galaxies discussed here have lower star-formation 
rates and classify as LBGs rather than SMGs or ULIRGs. The high-redshift 
galaxy ($z=7.5$) was detected with ALMA \citep{2015Natur.519..327W}. 
\citet{2017MNRAS.466..138K} find a current SFR of 13 \msunyr\ for 
a Chabrier IMF. The reported dust mass is $10{^{7.7\pm0.2}}$ \msun. 
The average star-formation rate over the past 80 Myr is $\sim$ 49 \msunyr\ 
for a Salpeter IMF \citep{2015Natur.519..327W}.

\citet{2017ApJ...837L..21L} observed a Y band dropout lensed by the cluster 
Abell 2744. The galaxy A2744$\_$YD4 has a spectroscopic redshift of 
$z=8.38$. ALMA observations indicate a dust mass of 
$5.5^{+19.6}_{-1.7}\times 10^6$ \msun\ and a star-formation rate of 
$20.4^{+17.6}_{-9.5}$ \msunyr\ for a Chabrier IMF.

Another Y band dropout Lyman-break galaxy, lensed by the cluster
MACS J0416.12403, was recently found to be at a redshift of 8.312 from detection
of the [\ion{O}{3}] 88 $\mu$m line in ALMA observations
\citep{2018arXiv180604132T}. The inferred dust mass is 
$3.6\pm 0.7\times 10^6$ \msun\ (for an assumed dust temperature of 50 K). The star 
formation rate is $\sim$ 14 \msunyr\  (for a Chabrier IMF) for an age of 180 Myr, with 
very large uncertainties, depending on the adopted extinction law.
\section{Results and discussion\label{s:discussion}}
\begin{figure}[htp]
\hspace*{-1.0cm}
\includegraphics[width=1.1\hsize]{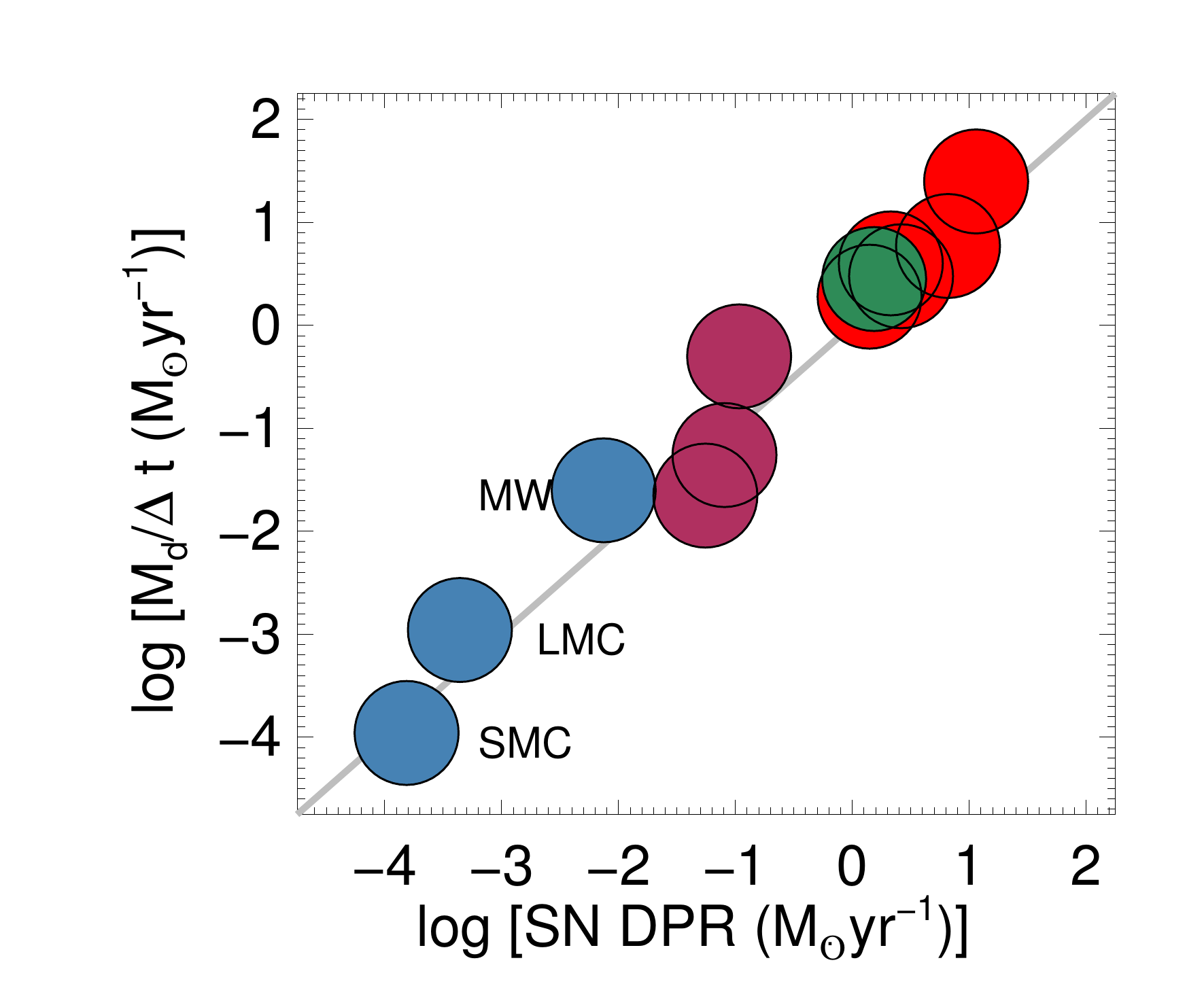}
\hspace*{-1.0cm}
\includegraphics[width=1.1\hsize]{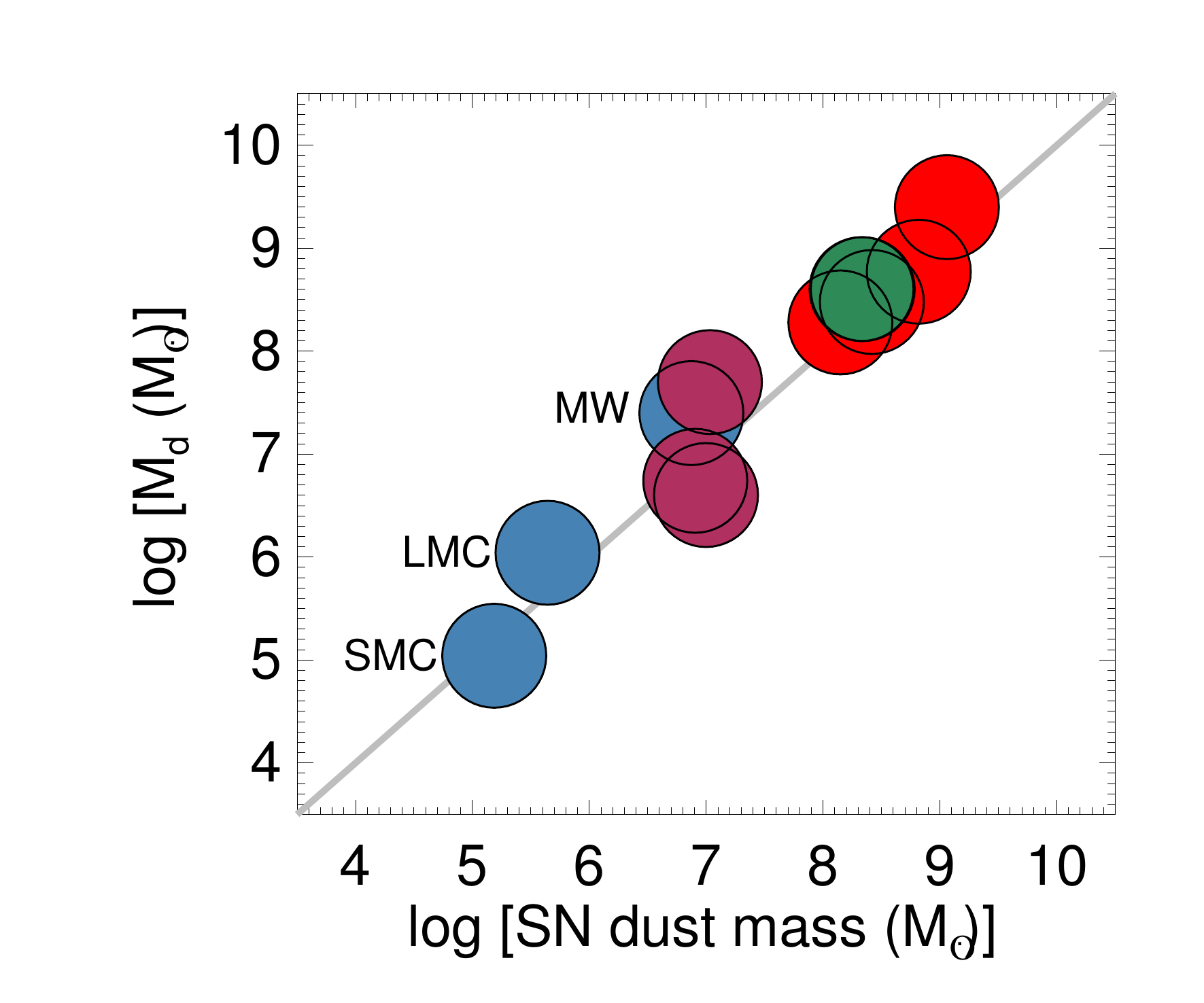}
\caption{
Observationally inferred values versus model predictions,
based on Equation (1) and Table~1. 
A value of $\mu_D=0.0022$ was adopted for consistency with the use 
of a Salpeter IMF to infer the observed star-formation rates. 
Uncertainties are assumed to be 0.4 dex 
(uncertainties in the DPR have been circularized for simplicity).
Blue circles signify Local Group galaxies, the green circle represents 
intermediate redshift SMGs ($z\sim 2$--3),
the red circles represent high-redshift SMGs ($z\sim 6$--7),
and the purple circles represent the highest redshift Lyman break 
galaxies ($z\sim 7.5$--8.4).
Lower panel: The model dust masses (Equation 1) are largely consistent with 
the observed dust masses over four orders of magnitude in dust mass and over 
the redshift range 0--8.4. 
Upper panel: The dust production rates (Equation 2) are consistent with the 
observed values over five orders of magnitude. 
}
\label{f:md}
\end{figure}
In Figure~\ref{f:md} (lower panel) we plot 
measured dust masses of galaxies against the total dust
mass produced by SNe, as estimated from Equation (1).
We also show the corresponding dust production 
rates inferred (Equation 2) in Figure~\ref{f:md} (upper panel). Remarkably, the predictions of the simple model are entirely consistent with the observed dust masses, within the considerable uncertainties over four orders of magnitude and across the redshift range 0--8.4, with dust production rates spanning five orders of magnitudes.
\subsection{Mass conservation constraint on maximally dusty galaxies}
\label{ss:max}
We have shown that observed dust masses are consistent with being produced by 
core-collapse SNe at levels similar to those detected in SN 1987A and the 
Crab, i.e., of order 0.3--0.4 \msun\ on average. In this sense, star-forming 
galaxies may be seen as maximally dusty: they contain the amount of dust SNe 
can produce, no more and no less. This is equivalent to using the `maximum 
efficiency' SN dust production models of \citet{2011A&ARv..19...43G}, as 
shown in Figure~\ref{f:eta}, which assume no subsequent destruction.

The consistency of the maximally dusty model with observed dust masses in 
star-forming galaxies is remarkable. While it is not proof that the dust 
is actually produced directly in SNe, it sets a strong constraint on 
models of dust formation in that other dust production mechanisms must obey 
this mass conservation constraint. For example, if dust is destroyed it must 
reform in about the same quantities.
\subsection{Reformation behind the reverse shock}
\label{ss:reformation}
The main concern in adopting SNe as the dominant dust factories in star-forming 
galaxies is that the dust could possibly be destroyed shortly after its formation (see 
Sections \ref{ss:snreverse} and \ref{ss:snforward}). Regarding destruction by the SN 
itself we now know that the dust apparently survives and even grows over the first 30 
years, as evidenced by the large amounts of ejecta dust in SN 1987A (see 
Fig.~\ref{f:devol}). In the coming years we will know if the reverse shock will eventually 
destroy the dust formed.

On the other hand we also note that while shocks may destroy dust, they might 
also induce either dust formation, re-formation or re-growth. A classical 
example where dust re-formation in a SN has been suggested is SN 2008S 
\citep{2011ApJ...743...73K}. Signatures pointing to either dust re-formation or re-
growth are seen behind the reverse shock in SN 1987A (M. Matsuura, private 
communication). Additionally, in Type IIn SNe, very early dust 
formation takes place in the cool dense shell, which is formed between the 
forward and reverse shock in the dense circumstellar medium. Prominent 
examples are SN 2005ip, 2006jd, 2010jl or 2011ht \citep[e.g.,][]
{2012ApJ...751...42S, 2012ApJ...760...93H, 2014Natur.511..326G}. 
A similar mechanism may foster 'in-situ' dust re-formation or re-growth 
behind a  reverse shock penetrating through ejecta dust, in which case 
the dust budget affected by a reverse shock would be a balance between
dust destruction and reformation. However, theoretical models do not find 
evidence for dust reformation in the post-reverse shock medium when assuming all 
dust has been destroyed by the reverse shock in the first place 
\citep{2014A&A...564A..25B}. Nevertheless, \citet{2014A&A...564A..25B} find 
that molecules reform and as discussed in Sections \ref{ss:snreverse} and 
\ref{ss:snforward}, whether or not dust gets destroyed may depend on the 
actual grain size distribution of the dust formed in a SN. 
%
\subsection{Large grains}
\label{ss:largegrains}
As discussed above, from a theoretical point of view, it has been argued 
that the majority of dust formed in SNe will get destroyed in a reverse 
shock \citep{2007MNRAS.378..973B, 2015MNRAS.451L..70M, 2016A&A...587A.157B} 
while others find a much smaller effect \citep{2007ApJ...666..955N}. 
The differences in the destruction efficiency of dust grains by SN reverse 
shocks can be traced back to the strong sensitivity to the markedly 
different initial grain size distributions, e.g., resulting from these models 
\citep{2001MNRAS.325..726T,2003ApJ...598..785N}. Similarly, variations in  
destruction time scales and efficiencies of ISM dust through SN shocks are 
also subject to the grain sizes considered (see Section \ref{ss:snforward}).    
Simulations and models of dust reprocessing of dust grains through shocks indicate 
that large fractions can survive \citep{2010ApJ...715.1575S}, especially if grains are 
large ($> 0.1 \mu$m) \citep{2004ApJ...614..796S, 2016A&A...589A.132B}.
Recent observations suggest the formation of a significant amount of large 
grains. From the extinction curve of dust formed in the Type IIn SN 2010jl,
\citet{2014Natur.511..326G} inferred the formation of large grains. 
Subsequently, this was also inferred from detailed modelling of the dust 
formed in SN 1987A and the Crab \citep{2015ApJ...801..141O, 
2015MNRAS.446.2089W, 2016MNRAS.456.1269B}. \citet{2018arXiv180104217B} 
also recently found that the extinction curve of dust formed in the Type IIn SN 
2005ip might be interpreted as being due to the formation of large grains. 
The formation of large grains in SNe is backed by theoretical models 
\citep{2015A&A...575A..95S} and by the observations of shallow extinction curves in strongly star-forming regions \citep{2016MNRAS.455.4373D}.
%
\subsection{Current state of affairs}
%
Under the assumption of efficient SN reverse shock dust destruction, one is driven to 
advocate accretion in the ISM to re-grow dust grains \citep{2009ASPC..414..453D, 2015A&A...577A..80M, 2015MNRAS.451L..70M} (AGB stars are not efficient dust 
producers on short time scales, i.e., at high redshift \citep{2011A&A...528A..14G}).
This process may be efficient in a MW environment, but
\citet{2015MNRAS.451L..70M} suggested from chemical evolution models that the 
accretion time scale in high-redshift galaxies must be about 200,000 yr. Such a short 
time scale\footnote{We note that in chemical evolution models (see Section 3), the 
entire evolution of a SN remnant is considered `instant'. On the other hand, as 
discussed in Section 2.4, the mixing of elements into the ISM may take many Myr.}
sets some stringent constraints on this scenario.

While the original suggestion was that accretion would occur in the cold neutral 
medium (CNM; $n\sim$10--100 cm$^{-3}$) \citep{2009ASPC..414..453D} it has been 
customary to assume that the main growth site in high-redshift galaxies is in molecular 
clouds at a much higher density ($>10^4$ cm$^{-3}$). However, as demonstrated by 
\citet{2016MNRAS.463L.112F} (see discussion in Section 2.4), grain growth cannot 
happen efficiently in molecular clouds of high density, hence one needs to reconsider 
the CNM \citep{2009ASPC..414..453D, 2016ApJ...831..147Z, 2016MNRAS.463L.112F}. In this case, the density is 2--3 orders of magnitude lower than in molecular 
clouds. As the accretion time scale is proportional to $a/n$ 
\citep{2013EP&S...65..213A,2016MNRAS.463L.112F}, where $a$ is the grain size, this requires grains 
that are smaller by 2--3 orders of magnitude, in order to provide the needed increased 
grain surface area. Therefore, nm-sized grains are required to be able to reach 
sufficiently short growth time scales at high redshift. Such small grains pose severe 
challenges \citep{2018ApJ...857...94Z}. In particular, the temperatures of such small 
grains will be so high, especially considering the high CMB temperature at high 
redshift, that accreted ions will not stick to the grains \citep{2016MNRAS.463L.112F}.

Alternatively, dust reformation and growth could happen on an even faster time scale, 
immediately after dust destruction, e.g., behind the reverse shock of the SN 
(Section 5.2). The constraint from our model is that the different scenarios (SN dust 
formation, reverse shock re-formation, or interstellar grain growth) must reach the 
same level of net dust production per SN. Given that interstellar destruction processes 
due to SN forward shocks or astration apply to any kind of dust, the need for limited 
destruction in high-redshift galaxies \citep{2011A&A...528A..14G, 2014ApJ...782L..23H} 
applies to any of these scenarios. This might favour processes which can produce 
large grains. 
\subsection{Summary}
\label{ss:summary}
%
We have gathered recent observational and theoretical results on SN dust 
formation and dust content in Local Group and high-redshift galaxies. The 
observed dust masses can be accounted for by a simple unifying model in which 
the majority of the dust is formed by SNe in the current star-formation 
episode. The model does not rule out other dust production scenarios but they 
are not strongly required either. It seems likely that if SNe produce the 
majority of the dust in massive high-redshift galaxies, they are also 
important sources of dust in Local Group galaxies.

We have not specified exactly how the dust must ultimately
be produced, such as, e.g., in the 
SN ejecta. The only assumption is that the dust is formed by SNe or massive 
stars on a short time scale, i.e., $\lesssim 10$--100 Myr. For example, 
significant dust could have been produced in mass loss events prior to the SN 
explosion (or fall-back black hole formation), or dust may have been formed 
in rapid grain growth scenarios out of the metals expelled by the massive 
star or SN. Rapid destruction and re-formation of the dust is of course also 
allowed, as long as the two terms are roughly balanced 
\citep{1998ApJ...501..643D}.

We conclude that the amount of dust in star-forming galaxies is consistent with
having been produced by SNe. Moreover, what has been destroyed must have
reformed rapidly. 
\acknowledgments
We thank Sebastian H\"onig and Mikako Matsuura for
helpful comments and suggestions.
C.G. acknowledges funding by the Carlsberg Foundation. 
J.H. was supported by a VILLUM FONDEN Investigator grant 
(project number 16599).
\clearpage
\bibliographystyle{yahapj}
\bibliography{dust}
\end{document}